\documentclass[12pt]{iopart}

\usepackage{graphics}

\begin{document}

\title{Foil-based atom chip for Bose-Einstein condensates}

\author{C. J. Vale\footnote[1]{To
whom correspondence should be addressed (vale@physics.uq.edu.au)}, B. Upcroft, M. J. Davis, N. R. Heckenberg
 and H. Rubinsztein-Dunlop
}                     

%
\address{Centre for Biophotonics and Laser Science,
School of Physical Sciences, University of Queensland,
Brisbane, Queensland 4072, Australia.}

\begin{abstract}
We describe a novel method of fabricating atom chips that are well
suited to the production and manipulation of atomic Bose-Einstein
condensates. Our chip was created using a silver foil and simple
micro-cutting techniques without the need for photolithography. It
can sustain larger currents than conventional chips, and is
compatible with the patterning of complex trapping potentials. A
near pure Bose-Einstein condensate of 4\,$\times$\,10$^4$ $^{87}$Rb atoms
has been created in a magnetic microtrap formed by currents
through wires on the chip. We have observed the fragmentation of
atom clouds in close proximity to the silver conductors. The
fragmentation has different characteristic features to those seen
with copper conductors.
\end{abstract}

\maketitle

\section{Introduction}
\label{intro} Bose-Einstein condensates (BECs) have become a
valuable tool for probing many aspects of atomic and quantum
physics.  Recently, condensates have been produced in miniature
magnetic traps formed by current-carrying wires patterned onto a
substrate \cite{ott01,hansel01}.  Such devices have become known
as atom chips, and are finding wide application in efforts to
coherently control matter for new atom-optical devices
\cite{folmanreview02}.

Atom chip BEC experiments have mostly been performed with chips
fabricated using photolithographic techniques
\cite{reichel01,fortagh02,leanhardt02}. These chips have been
highly successful and form the basis of a growing number
of experiments. However, there are some limitations to chips
produced in this manner.  One is that the photoresist thickness
and etching process constrain the maximum height of the conducting
wires. Typical conductor thicknesses range from about $2\,\mu$m
for evaporatively or sputter-coated wires, up to $10\,\mu$m for
electroplated wires. The latter also appear to suffer from rough
edges, leading to fragmentation of clouds near the surface
\cite{esteve04}. A desirable feature of atom chips is that
narrower wires result in tighter trapping potentials near the
surface for a fixed current.  However, the cross-sectional area
(and hence the thickness) of a wire restricts the maximum current
that can be sustained before it overheats and detaches
from the surface. Current densities of order $10^6$\,A\,cm$^{-2}$
have been achieved on atom chips at currents of order 1\,A, but
continuous operation at these current densities is fraught with
danger.  An elegant solution to this problem would be to use 
high-$T_c$ superconducting wires, although this introduces additional
technical complexity. Thicker extruded wire \cite{jones03} and
machined copper conductors \cite{schneider03} have also been used
to obtain higher currents, but these methods are not suited to
sub-millimetre patterning.

In this article we describe a method of fabricating atom
chips with thick wires capable of sustaining currents of several
amps without overheating. We begin by reviewing magnetic
microtraps and provide analytic expressions to describe the
properties of the widely used Z-wire trap.  We then describe the
fabrication process for our chip and demonstrate its effectiveness
by using it to produce a Bose-Einstein condensate of $^{87}$Rb.
Finally, we look at cold atom clouds trapped close to the chip surface and observe fragmentation of the atomic density profile.

\section{Magnetic Microtraps}

When an atom with a magnetic dipole moment, {\boldmath $\mu$}, is
placed in a magnetic field, {\boldmath $B$}, it experiences an
interaction potential $U$\,=\,-{\boldmath
$\mu$}\,$\cdot$\,{\boldmath $B$}.  If the projection of the
magnetic moment onto the field remains constant during the atom's
motion, the adiabatic condition is satisfied (i.e.\ the atom
remains in the same magnetic substate, $m_F$). In such cases the
potential is given by the scalar expression, $U$\,=\,-$m_F g_F
\mu_B B$, where $g_F$ is the Land\'{e} g-factor, $\mu_B$ is the
Bohr magneton and $B$\,=\,$|${\boldmath $B$}$|$.  Atoms whose
magnetic moment aligns antiparallel to {\boldmath $B$} are known
as weak-field seekers, as their lowest energy state is at a
minimum of $B$.

\begin{figure}\center

\resizebox{0.4\textwidth}{!}{%
  \includegraphics{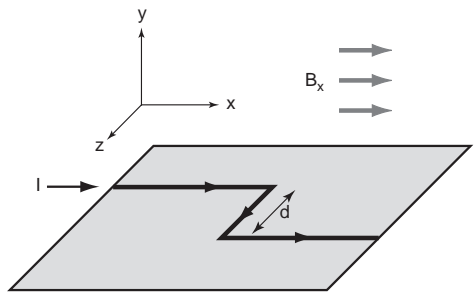}
}

\caption{A miniature Ioffe-Pritchard trap on an atom chip.  A
Z-shaped wire and homogeneous bias field can produce a simple trap
for weak-field seeking atoms above the centre of the middle
section of wire, as described in the text.}
\label{fig:zchip}       
\end{figure}

Several arrangements of wires have been proposed
\cite{weinstein95} and used in experiments to realise a variety of
magnetic trapping geometries for weak-field seeking atoms
\cite{hinds99,folmanreview02}.  A building block for all of these
is the ``side guide'', in which the combination of the field of a
straight conducting wire and homogeneous bias field produces a
two-dimensional, linear (to first order) trapping potential.  A
simple three-dimensional trap can be realised by bending a single
wire into the shape of a Z, as shown in figure \ref{fig:zchip},
where the top and bottom wire sections produce a confining
potential in the longitudinal ($z$) direction.

The combination of a Z-shaped wire and transverse bias field,
$B_x$, produces a Ioffe-Pritchard type trap with a nonzero minimum
located above the centre of the middle section of wire.  In the
$x$-$y$ plane the trap minimum is where the field of the wire is
exactly compensated by the bias field $B_x$. While the $x$ and $y$
components of the field go to zero here, the $z$-component, $B_z$,
is nonzero out of the plane of the chip ($y$\,=\,0), as the
fields due to the two sections of current in the $x$-direction add
constructively. Treating the wires as having an infinitesimal cross-section, the distance from the wire to the minimum is given
by
\begin{equation}\label{height}
    y_0 = \frac{\mu_0 I}{2 \pi B_x},
\end{equation}
where $\mu_0$ is the magnetic permeability of free space and $I$
is the current through the wire.  The magnitude of the field
gradient near the minimum ($x$,$y$,$z$)\,=\,(0,$y_0$,0) is equal
in the radial ($x$ and $y$) directions, and is given by
\begin{equation}\label{grad}
    \frac{\partial |B(0,y_0,0)|}{\partial x} = \frac{\partial
    |B(0,y_0,0)|}{\partial y} = \frac{\mu_0 I}{2 \pi y_0^2} =
    \frac{B_x}{y_0}.
\end{equation}
In the axial ($z$) direction the trapping potential is well
described by half of the $z$-component of field produced by two
infinite wires carrying current in the $x$-direction, located at
($y$,$z$)\,=\,(0,$\pm d/2$). The trapping potential is
given by
\begin{equation}\label{axtrap}
    B_{z}(0,y,z) = \frac{\mu_0 I}{4 \pi} \left ( \frac{1}
    {(z-d/2)^2+y^2} + \frac{1}{(z+d/2)^2+y^2} \right ),
\end{equation}
which is approximately harmonic at the centre with a curvature
\begin{equation}\label{curve}
    \frac{\partial^2 B_z}{\partial z^2} = \frac{16 \mu_0 I}{\pi}
    \frac{3d^2-4y^2}{d^2+4y^2}.
\end{equation}
The residual offset field in the $z$ direction at the minimum of
the trap is
\begin{equation}\label{offset}
    B_z(0,y_0,0) = \frac{\mu_0 I}{\pi}
    \frac{- 2 y_0}{d^2+4y_0^2},
\end{equation}
which can be partially compensated by an additional field
along -$z$ to increase the transverse confinement.

Near the minimum (0,$y_0$,0), the trapping potential can be
approximated as being ellipsoidal with the magnitude of the total
field
\begin{equation}\label{radtrap}
    |B(r,z)| = \left[B_z(0,y_0,z)^2 + \left( r \frac{\partial{|B_r|}}
    {\partial {r}} \right)^2\right]^{1/2},
\end{equation}
where $B_z(0,y_0,z)$ is given by equation~\ref{axtrap} and $\partial
|B_r| / \partial r$ by equation~\ref{grad}.

\section{Atom Chip}

\label{atomchip} 
\begin{figure}\center
\resizebox{0.4\textwidth}{!}{%
  \includegraphics{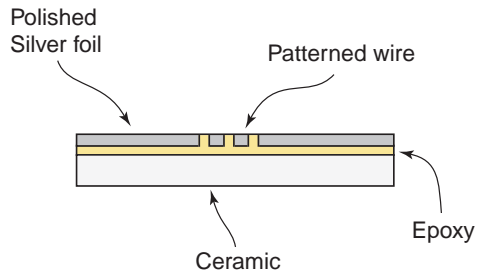}
} \caption{Schematic of the atom chip.  A silver foil was glued
onto a ceramic substrate using UHV compatible epoxy.  The polished
foil was approximately 90\,$\mu$m thick and the epoxy layer was
about half this thickness. Insulating channels were cut into the
foil with a micro-mechanical cutter and the channels were
back-filled with epoxy.}
\label{fig:chip}       
\end{figure}

Our atom chip was fabricated entirely in-house
using  materials and machinery readily available to typical
laboratories.  A schematic of the chip is shown in figure
\ref{fig:chip}.

A 125\,$\mu$m silver foil (Goodfellow AG000360) was glued onto a
1\,mm thick machinable ceramic (Macor) substrate using EpoTek H77
ultra high vacuum (UHV) compatible epoxy. Silver was chosen as the
conductive material as it has the lowest resistivity of all metals
(1.59\,$\mu \Omega$\,cm), compared with gold (2.2\,$\mu
\Omega$\,cm) and copper (1.67\,$\mu \Omega$\,cm).  It is also
highly reflective (97\%) to light resonant with the Rb D2
transition at 780\,nm. The epoxy adheres strongly to the ceramic,
but less effectively to the silver. Improved bonding to the silver
was achieved by roughening the surface to be glued with fine-grade
sand paper. A 100\,nm platinum layer was evaporatively coated onto
the foil, which was then oxidised (with heat in air) before
gluing. After curing the epoxy, the foil was polished to a mirror
finish in three stages. Firstly, fine-grade abrasive paper held on
a flat block was used under running water to take out any large
($>$\,10\,$\mu$m) ripples in the surface. Secondly, a range of
finer grade abrasive cloths (Micromesh 1500\,--\,8000) were used
down to a grain size of 3\,$\mu$m. Finally, 3\,$\mu$m and
1\,$\mu$m grain size diamond polishing pastes (ProSciTech M23/3
and M23/1) were applied to the surface with a silk polishing
cloth. These three polishing stages typically removed about
30\,--\,40\,$\mu$m of the silver, leaving a final conductor
thickness of approximately 90\,$\mu$m. The finished surface had
very few scratches visible to the naked eye. A near-infrared laser
beam was reflected from the finished surface and viewed several
metres away without visible distortion. The power in the
reflected beam was 95\,$\pm$\,2\% of the incident power, in good
agreement with the known reflectivity of silver at 780\,nm.

The polished chip was then ready to be machined.  A computer
numerically controlled (CNC) milling machine was used to cut
insulating gaps into the silver foil using a 150\,$\mu$m diameter
PCB cutting tool (LPKF 0.15\,mm end mill, 107244).  The wire
patterns were programmed into the CNC mill and the cutting was
performed in several runs, with the cut depth incremented by
20\,$\mu$m each run. The cutting tool spun at 10\,krpm and
traced the pattern of the wires at a speed of 5\,mm/min. We found
these parameters gave good results, but expect that higher rotation
rates would be desirable. Deeper cutting increments (up to
50\,$\mu$m) were also tried and no noticeable degradation of the
cuts were observed.  However, the tool was seen to wear very
quickly if it penetrated the ceramic and so we were cautious with
our increments. Small holes (1\,mm diameter) were drilled through
the foil, epoxy and macor  near the edges of the chip, in
the centre of the current connection pads.

After the pattern was cut, the resulting insulating channels were
carefully backfilled with epoxy to provide additional mechanical
support and heatsinking for the wires.  Finally, the whole chip
was repolished using the diamond pastes to remove any
scratches or residues from the cutting process.  The chip was
cleaned with distilled water and methanol, before the
electrical connections were assembled and the chip was placed in the vacuum chamber.

\begin{figure}\center
\resizebox{0.33\textwidth}{!}{%
  \includegraphics{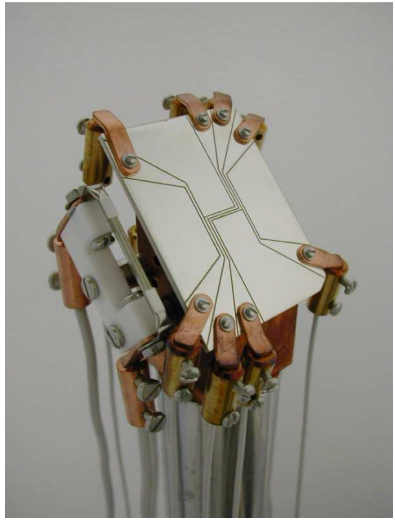}
} \caption{Photograph of the atom chip before being placed in the
vacuum chamber.  The electrical connections via the copper tabs
are visible on the four corners of the chips.  The rubidium
dispenser can be seen on the left side of the chip mount.}
\label{fig:chippic}       
\end{figure}

The robust nature of our chip allows simple yet secure electrical
connections to the wires.  The holes drilled through the
connection pads accommodate 1\,mm diameter screws. Copper tabs,
also with holes drilled through them, were placed onto the
connection pads and held in place by screws with a nut and sprung
washer that prevented loosening during bake-out of the vacuum
system. By contacting over a large area of the pad we ensure very
low resistance connections and negligible heating. Figure
\ref{fig:chippic} shows the completed chip with the electrical
connections in place.  The copper tabs are attached to barrel
connectors, which in turn connect to thicker ceramic coated wires
(Kurt J. Lesker, CCWA10SI and CCWA20SI) that run to the electrical
feedthroughs for the vacuum chamber. Also visible is the rubidium
dispenser, attached to the side and recessed by a few millimetres
from the silver surface of the chip.

We have tested the current-carrying capacity of our chip wires and
found that individual wires can sustain currents of 8\,A {\it in
vacuo} for 10\,s with the temperature remaining below
100\,$^{\circ}$C. The temperature was evaluated by monitoring the
resistance of the wire from outside the vacuum chamber. This
necessarily  includes the resistance of the connections and
connecting wires --- our estimate of the heating assumes that the
increase in resistance observed was entirely due to heating of the chip wires. We therefore obtain an upper limit for the
temperature increase of the chip. A nude ion gauge located
approximately 30\,cm from the chip could detect no increase of the
pressure in the vacuum chamber at the level of
2\,$\times$\,10$^{-11}$\, mbar during these measurements. The
current density in the wire was 5\,$\times$\,10$^4$\,A\,cm$^{-2}$.

Figure \ref{fig:chippic} also shows the wires patterned onto the
chip.  These consist of two parallel Z-trap wires with a centre to
centre separation of $2a$\,=\,0.4\,mm, and two additional endcap
wires for varying the axial confinement.  The parallel Z-wires are
capable of producing fields suitable for realising the two-wire
guide proposed in \cite{hinds01}. Assuming equal currents in the
two wires, there are three possible configurations in the $x$-$y$
plane.  These can be selected by the strength of the
(dimensionless) bias field, $\beta$\,=$B_x/B_0$, where
$B_0$\,=\,$\mu_0 I /2 \pi a$. The first, $\beta$\,$<$\,1, consists
of two quadrupole traps in the $x$\,=\,0 plane, separated
vertically by $2 a \sqrt{1-\beta^2} / \beta$. The second is a
single hexapole trap which occurs at the critical bias field value
($\beta$\,=\,1), with a minimum at ($x$,$y$)\,=\,(0,$a$).  The
third configuration is for $\beta$\,$>$\,1 and produces two
symmetric traps separated horizontally by $2 a \sqrt{\beta^2-1}/
\beta$.

In the axial direction there is only one trapping configuration,
as the atoms are much further away from the two end-cap wire
sections than the wire separation. The axial trap is well
described by equations~\ref{axtrap} and \ref{curve} for a single wire Z-trap carrying twice the current.

\section{Experiment}
\label{experiment}  We have used our atom chip to produce
Bose-Einstein condensates of $^{87}$Rb atoms. The experiment
follows the usual stages for BEC production and the details
of our procedure are described below.

Our vacuum chamber consists of a glass cell (made in our 
glass-blowing workshop) attached to a stainless steel vacuum chamber.  We use a
75\,l/s ion pump and a non-evaporable getter to reach pressures
below $2\,\times\,10^{-11}$\,mbar.  This was achieved by baking
the chamber for two weeks at a temperature of 140\,$^{\circ}$C and
pumping it with a turbo pump.  The epoxy bonding the silver to
the ceramic is rated to 150\,$^{\circ}$C which limits the
bake-out temperature.

Rubidium vapour is obtained from a dispenser source which is
pulsed with a current of 7.2\,A for 12\,s.  We collect 10$^8$
atoms in a reflection magneto-optical trap (MOT) formed 4.5\,mm
from the surface of the chip.  Trapping light is provided by a
Toptica DLX\,110 400\,mW external cavity diode laser system sent
through a single mode fibre. It is tuned 15\,MHz below the
5S$_{1/2}$ $F$\,=\,2 to 5P$_{3/2}$ $F'$\,=\,3 transition in
$^{87}$Rb. Four trapping beams enter the chamber, each with a
power of 23\,$\pm$\,1\,mW and 1/e$^2$ diameter of approximately
25\,mm. This laser is a home-built external cavity diode
laser system locked to the $F$\,=\,1 to $F'$\,=\,2 transition. The
repump laser is combined with the trapping laser on a polarising
beam splitter cube, and approximately 10\,mW of repump light
enters the chamber.

After the dispenser pulse, atoms are held in the MOT for a further
10\,s to allow the vacuum to recover. At this distance from the
surface the MOT does not appear to be seriously depleted by the
imperfections in our mirror due to the insulating gaps. While we
cannot make a direct comparison with a MOT with a near perfect
mirror, our atom number (10$^8$), temperature (90\,$\mu$K) and
lifetime (40\,s) are sufficient for producing condensates.

Next, the atoms are transferred to a compressed MOT with a magnetic
field provided by a U-shaped wire centred 1.6\,mm
below the silver surface carrying a current of 20\,A, combined
with a uniform transverse bias field of 12\,G. This 
results in  a spherical quadrupole field with gradients of
35\,G\,cm$^{-1}$ radially and 5\,G\,cm$^{-1}$ axially leading to a highly
anisotropic cloud at a distance of 1.5\,mm from the chip surface.
The MOT lasers and current through the U-wire are then turned off
simultaneously in less than 100\,$\mu$s and the atoms are
optically pumped into the $F$\,=\,2, $m_F$\,=\,2 ground state in
400\,$\mu$s. A current of 4\,A is then turned on through both
Z-wires on the chip and the atoms are caught in the potential
produced by the wire and a 12\,G transverse bias field. The bias
is ramped linearly from 12\,G to 30\,G over 500\,ms to compress
the cloud and achieve final trapping frequencies of 570\,Hz
radially and 11.2\,Hz axially at a distance of 430\,$\mu$m from
the surface. The lifetime of atoms in the magnetic trap is more than $30$\,s,
indicating that the vacuum is not adversely affected by the chip.
At this point efficient evaporative cooling can begin.

\begin{figure} \center 
\resizebox{0.7\textwidth}{!}{%
  \includegraphics{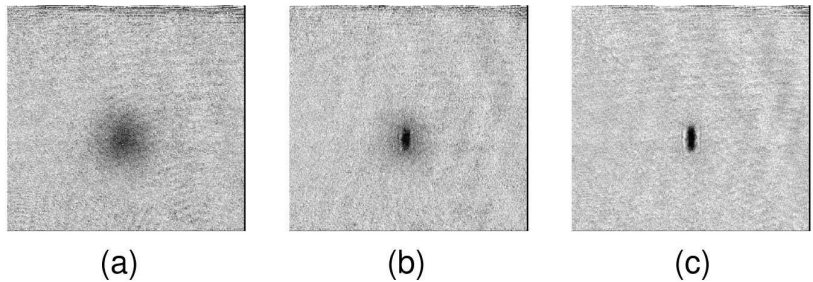}
}

\caption{Absorption images after 15\,ms free expansion of (a) a thermal cloud at 700\,nK, (b)
a partially condensed cloud at 450\,nK, and (c) an almost pure BEC
below 250\,nK. A condensate forms below
the critical temperature of about 500\,nK. The chip wires are visible at
the top of the images.}
\label{fig:bec}       
\end{figure}

The atomic cloud is evaporatively cooled through the BEC
transition using a single logarithmic sweep of a radio frequency
(RF) magnetic field from 13\,MHz to around 1\,MHz in 10.5\,s. In
the two-wire trap this produces condensates of around
3\,$\times$\,10$^4$ atoms. Larger condensates can be produced by
further compressing the cloud into a trap formed by current
through a single wire. Figure \ref{fig:bec} shows absorption
images of atom clouds following 15\,ms free expansion after being
released from a single wire magnetic trap. The three images show
clouds after terminating the RF evaporation at 1080\,kHz,
1050\,kHz and 1010\,kHz respectively. The trap forms 200\,$\mu$m
from the chip surface with oscillation frequencies of 1100\,Hz
radially and 6\,Hz axially.   The condensate contains around
5\,$\times$\,10$^{4}$ atoms and has a peak density of
4\,$\times$\,10$^{14}$\,cm$^{-3}$. With our parameters the
critical temperature for condensation is around 500\,nK, which is
reached with approximately 1.2\,$\times$\,10$^{5}$ atoms.

\section{Fragmentation}
When cold atom clouds are brought close to current-carrying wires,
the atomic distribution is seen to fragment into ``lumps''
\cite{kraft02,leanhardt03,jones04}. This is apparently due to
microscopic deviations in the direction of current flow which
produce small components of magnetic field parallel or
antiparallel to the axis of the wire \cite{kraft02}.  This effect
may prove to be a limitation of current-carrying wire-based atom
chips for applications such as atom interferometry.

\begin{figure} \center
\resizebox{0.7\textwidth}{!}{%
  \includegraphics{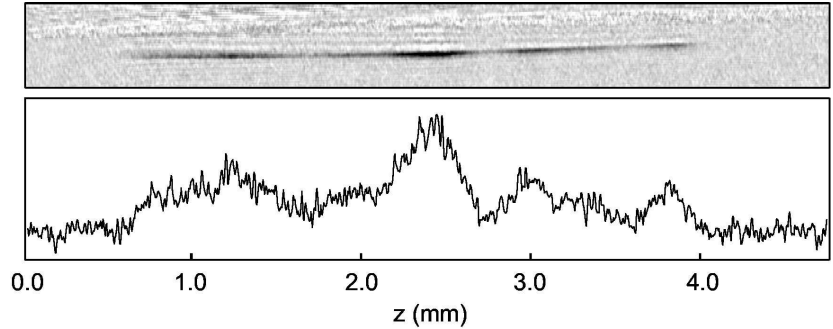}
}

\caption{Absorption image (top) and cross-sectional profile (bottom) of a
4\,$\mu$K fragmented atom cloud prepared at a distance of
45\,$\mu$m from the chip. The trap was accelerated away from the
surface for 5\,ms before imaging.}
\label{fig:frag}       
\end{figure}

We also observe fragmentation of cold clouds brought close to the
chip surface.  Figure \ref{fig:frag} shows an absorption image and
an averaged cross-sectional profile of a fragmented atom cloud
after being accelerated away from the chip surface. The cloud was
prepared at a distance of 45\,$\pm$\,5\,$\mu$m from a single wire
at a temperature of 4\,$\mu$K. 
At this distance the atoms could not be directly viewed in the
trap due to light scattered from the gaps between the wires.
Therefore the atoms were accelerated away from the surface for
5\,ms by ramping up the current through the wire immediately
before being imaged. Because of the short acceleration time and low
cloud temperature and axial trapping frequency (nominally
2.9\,Hz), any redistribution of atoms in the axial direction in
the displaced trap is not significant in the image.

We observe larger structure in the atomic density profile than has
been seen in other experiments --- up to the millimetre scale. The
cross-sectional profile also reveals shorter wavelength structure
superposed on these larger fragments.  This may be caused by
slight deviations in the movement of the cutter during milling
producing wires that are not perfectly straight ---
however we have no direct evidence of this as yet.  The depth of the fragmenting potential is comparable to what has been seen above copper conductors \cite{jones04}.

Fragmentation appears to be highly dependent on the geometric and
material properties of the conductors \cite{esteve04}.  Our method
of chip fabrication may be beneficial in this regard, as the 
high-purity solid metal foil should have superior conductor
uniformity to electroplated wires. It remains to be seen how
precisely wires can be patterned into a solid foil, and
subsequently how this affects fragmentation.  This topic will be
investigated in future work. Regardless of any improvements in
uniformity, our higher current-carrying capacity allows us to
produce moderately tight traps further away from the conductors.
This is important as fragmentation is seen to scale approximately
as $\rme^{-ky}/\sqrt{ky}$, where $k$ is the wavenumber of the
current deviations \cite{jones04}.

\section{Conclusion}
We have demonstrated a novel method of producing atom chips
suitable for the production and manipulation of Bose-Einstein
condensates. Our chip is capable of sustaining higher currents
than typical lithographically patterned chips, and can therefore
produce deeper magnetic traps. This facilitates condensate
production without sacrificing the versatility of the atom chip.
The patterning of wires on our chip is currently at the 100\,$\mu$m
scale but could be reduced using laser cutting to produce the
insulating channels. Nonetheless, our technique has proven to be
simple and reliable. Because of the higher current-carrying
capacity of this setup, we are able to produce moderately tight
traps at distances greater than 100\,$\mu$m from the surface,
where fragmentation effects become less important.

\ack
We are grateful to the Physics mechanical workshop at the University of Queensland, and in particular to Evan
Jones for excellent technical work in manufacturing the chip.  We
also thank J. Fort\'{a}gh and T. Campey for assistance with
experiments.  This work is supported by the Australian Research
Council.

\end{document}